\documentclass{article}

\usepackage{amsfonts}

\newcommand{\be}{\begin{eqnarray}}
\newcommand{\ee}{\end{eqnarray}}
\newcommand{\ben}{\begin{eqnarray*}}
\newcommand{\een}{\end{eqnarray*}}

\newcommand{\Tr}{\,{\rm Tr}\,}

\newcommand{\affiliation}[1]{}
\newcommand{\email}[1]{}
\newcommand{\keywords}[1]{}

\begin{document}

\title{Non-unique way to generalize the Boltzmann-Gibbs distribution}
\author{Jan Naudts\\
  Departement Natuurkunde, Universiteit Antwerpen UIA,\\
  Universiteitsplein 1, 2610 Antwerpen, Belgium
}

\email {Jan.Naudts@ua.ac.be}
\affiliation{
  Departement Natuurkunde, Universiteit Antwerpen UIA,\\
  Universiteitsplein 1, 2610 Antwerpen, Belgium
}

\date{February 2003}

\maketitle

\begin{abstract}
Alternative definitions are given of basic concepts of generalized
thermostatistics. In particular, generalizations of Shannon's
entropy, of the Boltzmann-Gibbs distribution, and of relative
entropy are considered. Particular choices made in Tsallis' nonextensive
thermostatistics are questioned.
\end{abstract}



\section{Introduction}


Several choices have to be made
when generalizing the Boltzmann-Gibbs distribution.
In this note
some of these choices are reviewed. Each time, two options are
presented: an A-option, which is the choice made in
Tsallis' thermostatistics \cite {TC88,CT91,TMP98,AMPP01},
and a B-choice which corresponds with generalized
thermostatistics as introduced by the author \cite {NJ02,NJ032}.
Advantages and disadvantages of each of the options are
discussed.

Starting point of the generalization is that the exponential
function appearing in the Boltzmann-Gibbs distribution
is replaced by some other increasing function. This idea
goes back to the early days of Tsallis' thermostatistics
\cite {TC94}. A generalization in this way is non-unique. As a consequence,
historically made choices may now be questioned. If they appear to
be suboptimal then the dilemma arises whether the standing
formalism should be modified. The present paper tries to start
this debate by clarifying possible choices and indicating
their consequences.

In the next section deformed logarithmic and exponential functions
are defined. Three sets of alternatives are presented in section 3.
In the final section some preliminary conclusions are drawn.

\section{Deformed logarithmic and exponential functions}

The point of view adapted in the present paper is to
replace the logarithmic and exponential functions by
arbitrary functions, which however share some of the properties
of the standard functions. In particular, following \cite {NJ02},
a deformed logarithmic function is denoted $\ln_\kappa(x)$.
It is defined for all positive $x$, and is an increasing and concave function.
It is normalized so that $\ln_\kappa(1)=0$. It could be further normalized
by requiring that
\ben
\int_1^0{\rm d}x\,\ln_\kappa(x)=1.
\een
However, some of the examples below do not satisfy this requirement.
Therefore, introduce the notation
\ben
F_\kappa(x)=\int_1^x{\rm d}y\,\ln_\kappa(y),
\een
and require only that $F_\kappa(0)$ is a finite number.

The inverse of the deformed logarithmic function $\ln_\kappa(x)$
is denoted $\exp_\kappa(x)$. Because the range of $\ln_\kappa(x)$
can be less than the whole real line, let us convene that
$\exp_\kappa(x)=0$ when $x$ is smaller than all values reached by $\ln_\kappa(y)$
and $\exp_\kappa(x)=+\infty$ when $x$ is larger than all values reached by $\ln_\kappa(y)$.

As an example, let us consider the definition of deformed logarithmic function
as it is used in the context of Tsallis' non-extensive thermostatistics.
It is denoted $\ln_q(x)$, where $q$ is a free parameter, which must lie between
0 and 2 in order for $\ln_q(x)$ to be a deformed logarithm according to the
definition given above. In what follows also an alternative definition will be needed.
The latter is denoted $\ln_q^{\bullet}(x)$. The two expressions are
\be
\ln_q(x)
&=&\frac{1}{1-q}(x^{1-q}-1)\cr
\ln_q^{\bullet}(x)
&=&\frac{q}{q-1}\left(x^{q-1}-1\right).
\label{tsallog}
\ee
A short calculation shows that
\ben
F_q(0)=\int_1^0{\rm d}x\,\ln_q(x)
&=&\frac{1}{2-q},\cr
F_q^\bullet(0)=\int_1^0{\rm d}x\,\ln_q^\bullet(x)
&=&1.
\een
Hence, the deformed logarithm $\ln_q(x)$ is not fully normalized.
The inverse functions are given by
\be
\exp_q(x)&=&\left[1+(1-q)x\right]_+^{1/(1-q)},\cr
\exp_q^\bullet(x)&=&\left[1+\frac{q-1}{q}x\right]_+^{1/(q-1)},
\label{tsalexp}
\ee
where $[x]_+$ equals $x$ when $x$ is positive, and zero otherwise.

Another example of deformed logarithmic and exponential functions
has been proposed by Kaniadakis \cite {KG01,KG02}.
Yet another example is found in \cite {NJ032}, where
it is used to describe the equilibrium distribution
of a single spin at the center of the Ising chain.
In what follows, only the definitions (\ref{tsallog}, \ref{tsalexp})
will be used to illustrate the impact of alternatives
in the context of Tsallis' thermostatistics.

\section{Three sets of alternatives}

\subsection{Entropy}

Let us start with Shannon's entropy functional, which for a discrete
probability distribution can be written as
\be
I(p)
&=&\sum_kp_k\ln\left(\frac{1}{p_k}\right)\cr
&=&-\sum_k\int_0^{p_k}{\rm d}x\,(1+\ln x).
\ee
Given these expressions, the two obvious generalizations are
\be
I_{\rm A}(p)&=&\sum_kp_k\ln_\kappa\left(\frac{1}{p_k}\right)\cr
I_{\rm B}(p)&=&-\sum_k\int_0^{p_k}{\rm d}x\,(F_\kappa(0)+\ln_\kappa x)\cr
&=&-\sum_k\int_0^1{\rm d}x\,p_k(F_\kappa(0)+\ln_\kappa(xp_k)).
\label{entdef}
\ee
Both definitions have all properties that one expects that
an entropy functional should possess.
An immediate advantage of $I_{\rm B}(p)$ over $I_{\rm A}(p)$
is that it is straightforward to calculate
derivatives. E.g., the derivative w.r.t.~temperature $T$ is given by
\be
\frac{{\rm d}I_{\rm B}(p)}{{\rm d}T}
=-\sum_k(F_\kappa(0)+\ln_\kappa(p_k))\frac{{\rm d}p_k}{{\rm d}T}.
\label{entdev}
\ee
The latter property is very convenient when proving thermodynamic stability
\cite {NJ032}.

Let us now consider how the definitions (\ref{entdef}) look like
in the special cases that $\ln_\kappa(x)$ is taken equal to
$\ln_q(x)$, respectively $\ln_q^\bullet(x)$, as given by (\ref{tsallog}).
One obtains
\ben
I_{\rm A}(p)
&=&\sum_kp_k\ln_q\left(\frac{1}{p_k}\right)\cr
&=&\frac{1}{1-q}\left(\sum_kp_k^q-1\right),\cr
I_{\rm B}(p)
&=&-\sum_k\int_0^{p_k}{\rm d}x\,(1+\ln_q^\bullet (x))\cr
&=&\frac{1}{1-q}\left(\sum_kp_k^q-1\right).
\een
Both definitions of entropy coincide.
The resulting expression is Tsallis' entropy \cite {TC88}.
It can also be written as (see formula (18) of \cite {TC99})
\ben
I_{\rm A}(p)=-\sum_kp_k^q\ln_q(p_k).
\een
In the latter form generalization to arbitrary deformed logarithmic functions
is not obvious.

For sake of completeness, the definitions of entropy  are given now for the case
of continuous distributions and for the quantum case.
If $\rho(\gamma)$ is a probability density
over some phase space $\Gamma$ then the expressions read
\be
I_{\rm A}(\rho)
&=&\int_\Gamma{\rm d}\gamma\,\rho(\gamma)
\ln_\kappa \left(\frac{1}{\rho(\gamma)}\right),\cr
I_{\rm B}(\rho)
&=&-\int_\Gamma{\rm d}\gamma\,\int_0^{\rho(\gamma)}{\rm d}x\,(F_\kappa(0)+\ln_\kappa x).
\ee
In the quantum case the entropy of a density matrix $\rho$ is given by
\be
I_{\rm A}(\rho)&=&\Tr \rho\ln_\kappa\left(\frac{1}{\rho}\right),\cr
I_{\rm B}(\rho)&=&-\int_0^{1}{\rm d}x\,\Tr\rho(F_\kappa(0)+\ln_\kappa(x\rho)).
\ee

\subsection{Canonical probability distributions}

Given discrete energy levels $H_k$, the Boltzmann-Gibbs distribution
equals
\be
p_k
&=&\frac{1}{Z(T)}\exp(-H_k/T)\cr
&=&\exp(G(T)-H_k/T).
\ee
In this expression $T>0$ is the temperature.
The normalization can be written either as a prefactor
$1/Z(T)$, or it can be included in the exponential as a term
$G(T)$. One clearly has $Z(T)=\exp(-G(T))$.
After generalization, the expressions become
\be
p_k^{\rm A}
&=&\frac{1}{Z(T)}\exp_\kappa(-H_k/T),\cr
p_k^{\rm B}
&=&\exp_\kappa(G(T)-H_k/T).
\label{eqpdf}
\ee
The expression for $p_k^{\rm A}$ has the advantage that
an explicit expression for the normalization exists
\ben
Z(T)=\sum_k\exp_\kappa(-H_k/T).
\een
The definition of $p_k^{\rm B}$ has the advantage that it
leads to an easy expression for $\ln_\kappa(p_k^{\rm B})$
\ben
\ln_\kappa(p_k^{\rm B})=G(T)-H_k/T.
\een
The latter is very convenient when calculating
the temperature derivative of the entropy.
Indeed, one obtains immediately, using (\ref{entdev}),
\be
\frac{{\rm d}I_{\rm B}(p^{\rm B})}{{\rm d}T}
&=&-\sum_k(1+\ln_\kappa(p^{\rm B}_k))\frac{{\rm d}p_k^{\rm B}}{{\rm d}T}\cr
&=&-\sum_k(1+G(T)-H_k/T)\frac{{\rm d}p_k^{\rm B}}{{\rm d}T}\cr
&=&\frac{1}{T}\sum_k H_k\frac{{\rm d}p_k^{\rm B}}{{\rm d}T}\cr
&=&\frac{1}{T}\frac{{\rm d}U}{{\rm d}T},
\label{ttcalc}
\ee
with the average energy $U$ given by
\be
U=\sum_kp_k^{\rm B}H_k.
\label{energdef}
\ee
Relation (\ref{ttcalc}) coincides with the thermodynamic definition of temperature
as the inverse of the derivative of entropy with respect to average
energy $U$.
\be
\frac{1}{T}=\frac{{\rm d}S}{{\rm d}U}.
\label{thermtemp}
\ee
This shows that $p^{\rm B}_k$ is the equilibrium probability distribution
of the canonical ensemble with entropy functional $I_{\rm B}(p)$
and with average energy $U$ defined in the usual way by (\ref{energdef}).
Moreover, the stability conditions, that $S$ is a concave function of $U$
and that $U$ is an increasing function of $T$, are satisfied.
Generically, the corresponding A-quantities do not have such nice properties.

In the general case, it is very difficult to write an explicit formula
expressing $p_k^{\rm A}$ in terms of $p_k^{\rm B}$. But this is feasible 
in the specific case that the deformed logarithm $\ln_\kappa(x)$ is given by
$\ln_q(x)$, respectively $\ln^\bullet_q(x)$, as defined by (\ref{tsallog}).
The expressions (\ref{eqpdf}) become
\be
p_k^{\rm A}
&=&\frac{1}{Z(T)}\exp_q(-H_k/T)\cr
&=&\frac{1}{Z(T)}\left[1-(1-q)H_k/T\right]_+^{1/(1-q)}\cr
p_k^{\rm B}
&=&\exp^\bullet_q(G(T)-H_k/T)\cr
&=&\left[1+\frac{q-1}{q}(G(T)-H_k/T)\right]_+^{1/(q-1)}.
\label{tsalpdf}
\ee
The first expression is the one introduced in \cite {CT91}.
The latter expression is found in \cite {NJ032}.
The two expressions look similar but differ in a number of aspects.
Let us try to match them.
Replace $q$ and $T$ in the former expression by
accented symbols $q'$ and $T'$. Then the two expressions
coincide provided that
\ben
q&=&2-q',\cr
G(T)&=&\frac{2-q'}{1-q'}\left(Z(T')^{q'-1}-1\right),\cr
T&=&T'Z(T')^{1-q'}.
\een
The latter expression makes clear that $p_k^{\rm A}$ and $p_k^{\rm B}$
have a completely different dependence on temperature $T$.
It is therefore obvious to check this temperature dependence
in existing applications of the probability distribution
$p_k^{\rm A}$. However, a first scan of the literature raises
the conjecture that this temperature dependence has not at all
been considered. There seems to be no evidence for temperature
dependent probability distributions of the form (\ref{tsalpdf}),
except of course in the $q=1$-case of the Boltzmann-Gibbs distribution.

It is straightforward to write down the extensions of (\ref{eqpdf})
in case of continuous distributions, or in the quantum case.

\subsection{Relative entropy}

The relative entropy of a discrete probability distribution $p$,
given a discrete probability distribution $r$, is defined by
\be
I(p||r)&=&\sum_kp_k\ln(p_k/r_k).
\label{relent}
\ee
It is only defined if $r_k=0$ implies that also $p_k=0$.
Relative entropy is also called Kullback-Leibler distance.
There are many ways to write (\ref{relent}), and hence,
many alternative definitions of generalized relative entropy.
Some possibilities are \cite {NJ03}
\be
I_{\rm A}(p||r)
&=&-\sum_kp_k\ln_\kappa\left(\frac{r_k}{p_k}\right),\cr
I_{\rm B1}(p||r)
&=&\sum_k\int^{p_k}_{r_k}{\rm d}x\,\ln_\kappa(x/r_k)\cr
&=&\sum_k\int^1_0{\rm d}x\,(p_k-r_k)\ln_\kappa\left(1+x\frac{p_k-r_k}{r_k}\right),\cr
I_{\rm B2}(p||r)
&=&\sum_k\int^{p_k}_{r_k}{\rm d}x\,(\ln_\kappa(x)-\ln_\kappa(r_k))\cr
&=&\sum_k\int^1_0{\rm d}x\,\left[
p_k\ln_\kappa(xp_k)-r_k\ln_\kappa(xr_k)
-(p_k-r_k)\ln_\kappa(r_k)\right].\cr
& &
\label{relentdefs}
\ee
The main advantage of $I_{\rm B2}(p||r)$ over the other expressions
is that, when $r$ equals the equilibrium distribution
$p^{\rm B}$, then one has (see \cite {NJ03})
\ben
I_{\rm B2}(p||p^B)
&=&I_{\rm B2}(p^{\rm B})-I_{\rm B}(p)
+\frac{1}{T}\sum_k(p_k-p_k^{\rm B})H_k.
\een
The quantity
\ben
\sum_kp_kH_k-TI_{\rm B}(p)
\een
is the free energy of the probability distribution $p$
at temperature $T$. The above result extends the standard result
that non-equilibrium free energy is a convex function which reaches
it minimum when $p$ equals the equilibrium distribution (in casu $p^{\rm B}$).
This is called the variational principle (see \cite {NJ03}).
The distance to the minimum, up to a factor $T$, is the relative
entropy. The other definitions of relative entropy do not have
such a property.

Let us now compare the different definitions in case
the deformed logarithms are given by (\ref{tsallog}).
One finds
\be
I_{\rm A}(p||r)
&=&-\sum_kp_k\ln_q(r_k/p_k)\cr
&=&\frac{1}{q-1}\sum_kp_k^q\left(r_k^{1-q}-p_k^{1-q}\right),\cr
I_{\rm B1}(p||r)
&=&\sum_k\int_{r_k}^{p_k}{\rm d}x\,\ln_q^\bullet(x/r_k)\cr
&=&\frac{1}{q-1}\sum_kp_k^q\left(r_k^{1-q}-p_k^{1-q}\right),\cr
I_{\rm B2}(p||r)
&=&\sum_k\int^{p_k}_{r_k}{\rm d}x\,(\ln_q^\bullet(x)-\ln_q^\bullet(r_k))\cr
&=&\frac{1}{q-1}\sum_k\left(p_k^q-r_k^q-qr_k^{q-1}(p_k-r_k)\right).
\label{tsalrelent}
\ee
The first two expressions coincide, but clearly differ from the last one.

The formulas for relative entropy in case of continuous distributions
are straightforward generalizations of the expressions (\ref{relentdefs}).
Quantization of relative entropy is not straightforward because two density
matrices $\rho$ and $\sigma$ are involved. When these do not
commute, then the order of operators is relevant.
For $I_{\rm A}(p||r)$ and $I_{\rm B1}(p||r)$
there is no obvious quantum equivalent, while for $I_{\rm B2}(p||r)$
a possible quantum expression is
\be
I_{\rm B2}(\rho||\sigma)&=&\int^1_0{\rm d}x\,
\Tr \left[
\rho\ln_\kappa(x\rho)-\sigma\ln_\kappa(x\sigma)
-(\rho-\sigma)\ln_\kappa(\sigma)
\right].
\ee

In the specific case that the deformed logarithms are of the form
(\ref{tsallog}) then it is clear how to write quantum generalizations
of all three definitions of relative entropy.
From (\ref{tsalrelent}) follows
\ben
I_{\rm A}(\rho||\sigma)=
I_{\rm B1}(\rho||\sigma)
&=&\frac{1}{q-1}\left(\Tr\rho^q\sigma^{1-q}-1\right)
=\frac{1}{1-q}\Tr\rho^q\left(\rho^{1-q}-\sigma^{1-q}\right),\cr
I_{\rm B2}(\rho||\sigma)
&=&\frac{1}{q-1}\Tr\left[\rho^q-\sigma^q-q\sigma^{q-1}(\rho-\sigma)\right].
\een
The former expression has been used in \cite {AS03}. The latter
expression is useful to prove a variational principle for the
quantum case.

\section{Discussion}

The present paper studies generalized thermostatistics from
the point of view that the exponential and logarithmic
functions appearing in the Gibbs formalism are replaced
by functions with similar properties. The obvious conclusion
is that there is quite some freedom in choosing generalizations.
Of course, there exist other points of view than the one presented here.
In particular, this paper avoids the question of extensivity
of macroscopic quantities like internal energy and entropy.
The Gibbs formalism behaves nicely under decomposition of
large systems into nearly independent subsystems. In
non-extensive thermostatistics a more complex
behavior is expected. The choices presented in this
paper have not been evaluated from this point of view.

It is also necessary to reanalyze existing applications of
Tsallis' thermostatistics with the intention to test
the different generalizations discussed in this paper.
Apparently, such tests of the basic assumptions
of generalized thermostatistics have been far from complete.


\end{document}